\documentstyle[prb,aps,preprint,psfig]{revtex}
\tighten
\begin{document}
\draft 
\title{Towards a quantum-chemical description of crystalline 
insulators: A Wannier-function-based Hartree-Fock study of Li$_{2}$O and 
Na$_{2}$O}
\author{Alok Shukla\cite{email}, Michael Dolg, and Peter Fulde} 
\address{Max-Planck-Institut f\"ur
Physik komplexer Systeme,     N\"othnitzer Stra{\ss}e 38 
D-01187 Dresden, Germany}
\author{Hermann Stoll} \address{Institut f\"ur Theoretische Chemie,
Universit\"at Stuttgart, D-70550 Stuttgart, Germany}

\maketitle

\begin{abstract}
A recently proposed approach for performing electronic-structure 
calculations on crystalline insulators in terms of localized orthogonal
orbitals is applied to the oxides of lithium and sodium, Li$_{2}$O and
Na$_{2}$O. Cohesive energies, lattice constants and bulk moduli of the 
aforementioned systems are determined at the Hartree-Fock level, and the
corresponding values are shown to be in excellent agreement with the
values obtained by a traditional Bloch-orbital-based Hartree-Fock approach.
The present Wannier-function-based approach is expected to be advantageous 
in the treatment of electron-correlation effects in an infinite solid by 
conventional quantum-chemical methods.
\end{abstract}
\section{INTRODUCTION}
\label{intro}
A typical quantum-chemical investigation of a system, employing a
wave-function-based approach, begins with a Hartree-Fock (HF) calculation which
provides an initial mean-field description of the system. Then, if the
need arises, it is improved systematically ---  the influence
of electron correlations is included by considering virtual excitations from the
HF wave function. However, in such a scheme the computational effort
has a rather unfavorable dependence on the size of the system and in the
best-case scenario it scales roughly as 
$N^{5}$, where $N$ is the number of atoms in the 
system. (Only recently, attempts towards achieving linear scaling at the HF
level~\cite{on-scf}, and towards exploiting the local nature of correlation effects
are being made~\cite{on-corr}.) Because of this proliferation in the computational effort, the
field of quantum chemistry is generally regarded as a science of finite 
systems with most of its applications limited to small or medium-sized 
molecules. Therefore, a naive extension of this quantum-chemical scheme
to treat a system such as a three-dimensional crystalline
solid is bound to run into problems. 
On the one hand a crystalline solid, for all practical
purposes, is an infinite system. It essentially has an infinite number
of electrons as well as an infinite spatial extension compared to the
dimensions of a unit cell.  
On the other hand, the translational symmetry force its orbitals   
to have a delocalized itinerant character. Clearly, with such crystal orbitals,
known as Bloch orbitals, it can be problematic to sum the
contributions of the electron repulsion part of the Hamiltonian to convergence.
In Bloch-orbital-based theories, the problems of infinite size and
delocalized orbitals have been overcome at the 
Hartree-Fock level, nowadays, by adopting suitable procedures of integration
over the Brillouin zone and lattice sums, as is done, for example, in
the program CRYSTAL~\cite{crystalprog}. However, inclusion of
electron-correlation effects by going beyond the Hartree-Fock level
within a traditional quantum-chemical scheme (such as the 
configuration-interaction or coupled-cluster approaches)
appears to be fraught with problems
if one uses Bloch orbitals as a basis set.
Complications will arise both due to the itinerant nature of the
Bloch orbitals, as well as the infinite number of virtual states spanning the
Brillouin zone that one will have to take into account, in order to
compute quantities such as the correlation energy. 

 If one abandons the built-in translational symmetry associated with the
Bloch orbitals, one can alternatively describe the electrons in a solid
as being localized entities associated with the atoms or bonds constituting
it. Such a localized description of electrons in a solid is much easier
to understand for a chemist who is usually thinking in terms of bonds. 
Unlike the Bloch-orbital-based 
approach, a localized-orbital-oriented description of a system is not 
unambiguous, e.g.\ in the sense that one can choose either an orthogonal or a 
nonorthogonal set of localized orbitals. 
Nonorthogonal orbitals, in general, are better localized and, in principle,
are also capable of describing the metallic systems where the electrons
are indeed quite itinerant. Recently 
Bella\"{\i}che and L\'{e}vy~\cite{levy} have presented an approach, along with
its application to crystalline LiH, which allows the determination of
{\em nonorthogonal} localized Hartree-Fock orbitals of an infinite solid,
from a set of finite-cluster calculations.
However, with nonorthogonal orbitals a 
conventional quantum-chemical treatment of electron correlation such as the 
one based on the configuration-interaction approach is considerably more 
complicated. On the other hand, the orthogonal orbitals, although less 
localized than their nonorthogonal counterparts, offer the possibility
of straightforward inclusion of correlation effects. Such an alternative
approach to the description of crystalline electrons in terms of localized
orthogonal orbitals was proposed by Wannier long ago~\cite{wannier}, and
therefore, such orbitals in the condensed-matter community have come to
be known as Wannier functions. It should be intuitively obvious that
it will be very difficult to describe metallic electrons in terms
of Wannier functions because the orthogonality requirement will
force the orbitals  to extend far away from their centers,
leading to their eventual delocalization~\cite{kohn}. 
However, for systems with a band gap such as semiconductors and insulators 
it should be possible, in principle, to satisfy both the orthogonality as 
well as the localization requirements as embodied in Wannier 
functions~\cite{kohn}.

Therefore, it should be clear from the preceding discussion, that if one
wants to study the influence of electron correlations on an infinite solid
using a traditional wave-function-based approach, it is mandatory that one 
first obtains a HF representation of the system in terms of Wannier functions.
This goal can be achieved by first performing a HF calculation for the
infinite crystal using a Bloch-orbital-based approach, and then localizing 
the Bloch orbitals using one of the many available localization 
schemes~\cite{local} to obtain the Wannier functions~\cite{mazari}. 
However, in our previous
papers~\cite{shukla1,shukla2} we had presented an alternative scheme which
allows a direct determination of Hartree-Fock Wannier functions of 
a crystalline insulator. In the first paper (henceforth to be referred to as
I), we had presented an outline of the formalism and used it to compute
the total energy per unit cell of crystalline LiH within an LCAO scheme.
In the second paper (to be called II), we gave a rigorous derivation of
the approach by minimizing the total energy of an infinite crystal, and
demonstrated its theoretical equivalence to a traditional Bloch-orbital-based 
HF approach. Additionally in II we applied the approach to obtain the
HF Wannier functions of LiF and LiCl crystals which were subsequently used to
compute the X-ray structure factors and directional Compton profiles.
In all the cases it was demonstrated that our results were in excellent
agreement with equivalent calculations performed using the CRYSTAL program,
which employs a Bloch-orbital-based HF approach. All the details pertaining
to the computer implementation of our LCAO-HF program were also
presented in II~\cite{shukla2}. Recently, we extended this approach
to study the band structure of insulating crystals
and applied it to the case of NaCl~\cite{albrecht}. 
We feel, however, that it is important
for us to study the efficacy of the present approach in a variety of
insulating systems. With that goal in mind, in the present paper, we
present the results of our Wannier-function-based HF calculations on
crystalline Li$_{2}$O and Na$_{2}$O. Both these crystals are ionic in nature
with the valence Wannier functions residing on O$^{--}$. However,
O$^{--}$ ion is not stable in its free form and becomes stabilized in the
solid state only due to the crystal field. Therefore, we believe that 
obtaining the valence Wannier functions of compounds involving O$^{--}$ is a 
nontrivial application of our approach.  Moreover, this application involves
rather diffuse basis functions requiring a significant improvement of the
accuracy of lattice sums with respect to our calculations described in II.

The remainder of our paper is organized as follows.
In section \ref{theory}, we briefly review the theoretical formalism.
Then in section \ref{results} we present the results of our calculations
for the Li$_{2}$O and Na$_{2}$O 
crystals. These results include cohesive energies, lattice constants
and bulk moduli of the two compounds at the Hartree-Fock level.
Our results are also shown to be in excellent agreement with those obtained 
using the CRYSTAL program. 
Finally, in section \ref{conclusion} we present our conclusions. 

\section{THEORY}
\label{theory}
In this section we briefly review the theory behind our approach.
For more detailed derivations we refer the reader to papers I and
II~\cite{shukla1,shukla2}. The essential idea behind the approach is
that in order to describe a perfect crystalline insulator in terms
of Wannier functions, one only needs
to obtain the Wannier-type orbitals localized in any one of its unit 
cells---the orbitals localized in the rest of the unit cells being their 
translated 
copies. This fact is a consequence of translation symmetry 
and can be stated mathematically as 
\begin{equation}
|\alpha({\bf R}_{i})\rangle = {\cal T} ({\bf R}_{i}) 
|\alpha({\bf 0})\rangle \mbox{,}
\label{eq-trsym}
\end{equation}
where $|\alpha({\bf 0})\rangle$ represents a Wannier orbital localized in
the reference unit cell assumed to be located at the origin while  $|\alpha({\bf R}_{i})\rangle$ is the corresponding orbital
of the $i$-th unit cell located at position ${\bf R}_{i}$, 
where ${\bf R}_{i}$ is a lattice vector. The corresponding translation
is induced by the operator ${\cal T} ({\bf R}_{i})$. Clearly, in a 
Wannier-function-based description, the entire crystal can be described
by the orbital set $\{ |\alpha({\bf R}_{j})\rangle; \alpha =1,n_{c}; j=1,N \}$, 
where $n_{c}$ is the number of Wannier functions associated with a unit cell
and $N$ is the total number of unit cells in the crystal whose limiting value,
of course, is infinity. In the following discussion Greek labels $\alpha$,
$\beta, \ldots,$ will be reserved for labeling the (occupied) Wannier orbitals of a 
unit cell.  

Having established the fact that one needs to specify  the Wannier orbitals
of only one of the unit cells, we need a dynamic prescription for obtaining 
them.  In paper I~\cite{shukla1} we adopted an 
``embedded-cluster'' picture of an infinite solid to achieve that goal.
In this scheme we envision the infinite crystal as a 
central cluster(the reference cell) embedded in the field of its environment 
consisting of an infinite number of self-similar unit cells arranged 
according to the crystal lattice. Within this picture,
it is straightforward, for a {\em closed-shell} system, to write down  the restricted 
Hartree-Fock (RHF) equations for the 
Wannier functions of the embedded cluster,  under the Born-Oppenheimer 
approximation:
\begin{equation}
( T + U
 +   \sum_{\beta} (2 J_{\beta}-  K_{\beta})   
+\sum_{k \in{\cal N}} \sum_{\gamma} \lambda_{\gamma}^{k} 
|\gamma({\bf R}_{k})\rangle
\langle\gamma({\bf R}_{k})| ) |\alpha\rangle
 = \epsilon_{\alpha} |\alpha\rangle
\mbox{,}
\label{eq-hff1}         
\end{equation}  
where $|\alpha\rangle$ stands for
$|\alpha({\bf 0})\rangle$, an orbital centered in the reference unit cell,
$T$ represents the kinetic-energy operator, $U$ represents
the interaction of the electrons of the reference cell with the nuclei
of the whole of the solid, while $J_{\beta}$, $K_{\beta}$ are the conventional Coulomb and exchange 
operators defined as 
\begin{equation}
\left.
 \begin{array}{lll}
 J_{\beta}|\alpha\rangle & = & \sum_{j} \langle\beta({\bf R}_{j})|\frac{1}{r_{12}}|
\beta({\bf R}_{j})\rangle|\alpha\rangle \\  
 K_{\beta}|\alpha\rangle & = & \sum_{j} \langle\beta({\bf R}_{j})|\frac{1}{r_{12}}|\alpha\rangle
|
\beta({\bf R}_{j})\rangle \\  
\end{array}
 \right\}  \mbox{.} \label{eq-jk} 
\end{equation}
Although here we have outlined an intuitive derivation of Eq.(\ref{eq-hff1}),
it can also be obtained rigorously by minimizing the total energy of the
infinite solid as shown in II~\cite{shukla2}.
The first three  terms on the left hand side of Eq.(\ref{eq-hff1}) 
constitute the conventional canonical Hartree-Fock operator while the 
last term plays the role of a localization potential.
It involves a sum over projection
operators constructed from the orbitals of the unit cells localized in the
immediate neighborhood ${\cal N}$ of the central cluster. 
For an infinitely high value of the shift parameters, $\lambda_{\gamma}^{k}$, it 
has a localizing effect on the orbitals of the reference cell
while simultaneously making the latter orthogonal to those located in ${\cal N}$.
The choice of ${\cal N}$ 
will clearly be dictated by the localization characteristics of the electrons
of the system under consideration. It is expected to be larger for systems
with smaller band gaps.
 In our calculations we have typically chosen ${\cal N}$ to include
up to third nearest-neighbor unit cells of the reference cell. 
The orthogonality of the orbitals contained in unit cells beyond ${\cal N}$
will also be dependent upon the band gap of the system and should be 
automatic once the region ${\cal N}$
has been chosen to be sufficiently large. 
Numerical values in the range $10^{3}$---$10^{5}$ atomic units  for the 
shift parameters $\lambda_{\gamma}^{k}$ were found to be suitable.

We have adopted a linear-combination of atomic orbital (LCAO) formalism
utilizing lobe-type Gaussian basis functions~\cite{whitten} to solve 
Eq.(\ref{eq-hff1}). Terms $U$, $J$ and $K$ appearing in 
Eq.(\ref{eq-hff1}) involve infinite
lattice sums and deserve special consideration. Evaluation of these
terms along with other computational aspects are discussed in detail in
paper II~\cite{shukla2}. 
It is clear that the orthogonalization of the orbitals of the reference
cell to those of the region ${\cal N}$ will introduce oscillations in these orbitals which are also referred to
as the orthogonalization tails. In order to describe these orthogonalization
tails, one needs to express the orbitals of the reference cell as linear 
combination of basis functions located both in the reference cell as well
as in ${\cal N}$. This increases the dimension of the Fock matrix
to be diagonalized as compared to a canonical Hartree-Fock program
as implemented, e.g., in CRYSTAL~\cite{crystalprog}. This, however, does not 
affect the evaluation of one- and two-electron integrals (and their number) if a careful use of 
translational symmetry is made as discussed in II~\cite{shukla2}.
After the diagonalization of the Fock matrix, one is
confronted with the task of choosing the occupied orbitals from the spectrum 
of the eigenvalues. For the ionic systems considered here, the aufbau principle
was used for this purpose; in practice, one starts with a suitable guess
for the Wannier functions, and iteratively solves Eq.(\ref{eq-hff1}) until
the energy per unit cell has converged.
 
\section{CALCULATIONS AND RESULTS}
\label{results}
In this section we present the results of the calculations performed on
crystalline Li$_2$O and Na$_2$O. These compounds have been investigated
extensively at the Hartree-Fock level in a number of papers by the Torino
group using their  CRYSTAL program~\cite{torino0,torino1}. Therefore, we
also intend to compare our results with the most recent results of that 
group~\cite{torino1}.
In addition, we discuss the characteristics of the O$^{2-}$ Wannier
orbitals obtained from our calculations and compare them to atomic orbitals
of free oxygen (and its singly negative ion).

\subsection{Geometry, basis set and computational parameters}
All the calculations reported below have been carried out by assuming
the observed anti-fluorite structure with the space group $Fm3m$.  
The reference unit cell was assumed to be the primitive cell with oxygen anion
placed at the origin and the two cations at the positions $(a/4,a/4,a/4)$ and
$(-a/4,-a/4,-a/4)$, where $a$ is the lattice constant. 

The Torino group studied these compounds using 
highly-optimized extended basis sets and for the sake of 
comparison, we have used the most 
recent basis set reported by them~\cite{torino1}.
It employs a (7s1p)/[2s1p] set for lithium, a (15s7p)/[4s3p]
set for sodium and a (14s6p)/[4s3p]
set for oxygen~\cite{torino1}.  As 
indicated in the previous section, we have not used real Cartesian-type  
Gaussian basis functions but rather their approximate counterparts obtained by 
forming linear combinations of lobe-type (1s) 
Gaussian functions~\cite{shukla2,whitten}. For the case of Li$_2$O the
number of (contracted) basis functions per unit cell is 23 while for the case of 
 Na$_2$O it is 39.
However, in order to satisfy the orthogonality requirements associated with the Wannier
functions, we also  have to supply the basis functions located on the 
neighboring cells to describe the Wannier orbitals of the central cluster.
As mentioned in the previous section, this neighborhood ${\cal N}$
consists of up to third-nearest neighbors of the central cluster. Since for
the fcc geometry, there are 42 unit cells in ${\cal N}$ so defined, one
has to use the basis functions of 43 unit cells including the reference
cell. Thus the number of basis functions associated with the Wannier orbitals
of Li$_2$O and Na$_2$O explodes to 989 and 1677 respectively.
This proliferation in the number of basis functions does not affect the
integral evaluation time in any way~\cite{shukla2}, compared to the CRYSTAL
program~\cite{crystalprog}, as long as the use of point-group symmetry
is the same. However, it does increase the
dimension of the Fock matrix to be diagonalized to the corresponding
numbers, thus increasing the time needed to perform the Hartree-Fock
iterations. The latter problem is not too critical, though, since contracting
basis functions in ${\cal N}$ to about single-zeta quality is expected to
remedy this issue without adversely affecting the accuracy of the calculation.

The CRYSTAL program uses several 
computational parameters which determine the accuracy of the Coulomb and the 
exchange series. The parameter related to the Coulomb series is called
ITOL1 and those related to the exchange series are called ITOL3, ITOL4 and
ITOL5~\cite{crystalprog}. The values of 7,7,7 and 15 
for these parameters for the CRYSTAL-program based calculations are 
generally believed to lead to well-converged results~\cite{crystalprog}.
These values of the parameters ensure an
absolute accuracy of $\approx 1.0\times10^{-7}$ atomic units (a.u.) in the
Fock matrix elements leading to an expected accuracy of $\approx$ 1 
milliHartree per atom in the total energy~\cite{crystalprog}.
Therefore, to make the comparison with CRYSTAL results transparent, in our
calculations also we treated the Coulomb and the exchange series in an
entirely equivalent way.  

\subsection{Lattice Constant, Bulk Modulus and the Cohesive Energy}
In order to optimize the lattice constant and to obtain the bulk modulus we 
first computed total energies per unit cell of both the compounds for 
different values 
of lattice constants which are closely spaced around the equilibrium value.
As also reported in papers I and II~\cite{shukla1,shukla2}, the total energies (per unit
cell) obtained by us
at different lattice constants agreed with
the corresponding values obtained by the CRYSTAL program to within fractions of
a milliHartree.
We then fitted these data points to polynomials.
The resulting bulk properties were found to be stable to within a fraction of a percent
with respect to an increase in the degree of the polynomial considered, and
are listed in table \ref{tab-eqprop}.
In the same table, lattice
constants, bulk moduli, and cohesive 
energies for the two compounds obtained from our calculations are compared
with the results of the Torino group~\cite{torino1} and also
with experiments whenever possible.  Cohesive energies were obtained by
subtracting the corresponding atomic Hartree-Fock energies from the
equilibrium total energy per unit cell. To make the comparison with
the results of the Torino group meaningful, we used the same atomic HF 
energies of -7.4313 a.u. (Li atom), -161.8513 a.u. (Na atom) and 
-74.8012 a.u. (O atom) as used in their calculations~\cite{torino1}. 
 
With the same basis sets, our approach and that of the Torino group~\cite{torino1}
should yield identical results.
A quick glance at table \ref{tab-eqprop} reveals that  the
agreement for lattice constants and cohesive energies is excellent. The 
maximum disagreement of 1 milliHartree in the cohesive energies is well
within the expected numerical accuracies of CRYSTAL~\cite{crystalprog} and
our program. The maximum deviation in the bulk moduli is less than 3 GPa
and are to be expected for the present disagreements in total energies,
since second derivatives are much more sensitive to numerical inaccuracies.
We suspect that part of the reason behind these small differences in the results
of the two programs could also be due to our use of lobe functions to approximate the
Cartesian-type Gaussian basis functions used in the 
CRYSTAL program~\cite{crystalprog}.

The experimental value of the lattice constant of Li$_2$O is based on 
inelastic neutron scattering experiments~\cite{expali2o} performed
in a temperature range 293-1603 K. The value of 4.573 $\AA$ is the 
zero-temperature value obtained by extrapolating the $a$ versus $T$ 
curve~\cite{expali2o}.
The result of the Torino group~\cite{torino1} is in essentially exact agreement with
this value while our value is 0.003 $\AA$ shorter. For Na$_2$O
the experimental value of the lattice constant was determined to be
5.55 $\AA$  at room temperature by Zintl et al.~\cite{expana2o} 
in an old experiment based upon powder diffraction pattern data.  
Because of the lack of availability of any other measurements, 
Dovesi et al.~\cite{torino1} extrapolated this to a zero-temperature
value of 5.49 $\AA$ using arguments based upon trends observed in the
temperature dependence of the lattice constant of NaF. Our calculated
value of $5.481$ $\AA$ as well as the value of $5.484$ $\AA$ of
the Torino group ~\cite{torino1} are both in good agreement with the
above-mentioned zero-temperature value. 
 
The extrapolated zero-temperature value of the bulk modulus of 
Li$_2$O obtained from the finite temperature data of Hull et 
al.~\cite{expali2o} is 89 GPa. Our value of 94.6 GPa is
approximately 6\% larger than this experimental value.
We believe that
most of this disagreement is due to missing correlation
effects. In the case of Na$_2$O, no experimental value is
available for the bulk modulus, to the best of our knowledge. However, reasonably
close agreement with the results of the Torino group~\cite{torino1}, gives us 
confidence as to the correctness of our result. 

When we compare the Hartree-Fock cohesive energies reported here to the 
experimental values~\cite{beexp}, we note that for Li$_2$O we recover
$\approx$ 67\% of the total contribution while for Na$_2$O this fraction
is about 57\%. As also argued
by Dovesi et al.~\cite{torino1}, most of the missing cohesive energy at
the Hartree-Fock level is due to the absence of correlation effects.
This belief is also substantiated by various finite-cluster-based calculations
performed in our group where, by including correlation effects, typically about
95\% of the experimental cohesive energy was recovered~\cite{corr}. 
Therefore, the extension of the scheme of local correlation-energy increments 
as described in
refs. \cite{corr}, to the case of infinite crystals, appears to be 
worthwhile. 
This task, however, is nontrivial and presently we are in the
process of implementing it.
\subsection{Wannier Functions}
\label{sec-wan}
Now we turn to the discussion of the Hartree-Fock Wannier functions that we
obtain on solving Eq.(\ref{eq-hff1}). Since the qualitative behavior of the 
Wannier functions of both compounds is expected to be the same, we will 
discuss only the orbitals of Li$_2$O.   
The central cluster chosen in this work consists of one primitive cell 
of the lattice. Therefore, for Li$_2$O our Hartree-Fock equations involve 
fourteen electrons leading to seven Wannier functions.
Out of these seven functions, three correspond to low-lying 
1s-type core orbitals
of the three atoms in the unit cell, while the remaining four correspond to
the 2s- and 2p-type functions centered on oxygen. Since these four 
high-lying valence Wannier functions are responsible for most of the 
chemical properties of the compounds considered, we will restrict our 
discussion to them.

The nearest-neighbor environment of the O$^{--}$ ion in the crystal
consists of Li$^+$ ions located at the eight corners of a cube 
with coordinates $(\pm a/4, \pm a/4, \pm a/4)$, while the center
of the cube at  position (0,0,0) is occupied by O$^{--}$ ion itself.
This type of environment results in three 2p-type Wannier functions of
the crystal, each one of which is a mixture of the 2p$_{x}$, 2p$_{y}$
and 2p$_{z}$ type basis functions localized on oxygen. 
In Figs. \ref{fig-comp2s111} and \ref{fig-2p111}, respectively, we present 
the 2s-type and one of the 2p-type Wannier functions of Li$_2$O crystal, 
localized on the O$^{--}$ ion. The 2p-type Wannier function presented
here is isotropically oriented along the body diagonals.
Both the orbitals are plotted along the
crystal [111] direction.
Since the crystal structure of Li$_2$O is invariant under the parity operation,
we expect the corresponding 2s- and 2p-type Wannier functions to be,
respectively, symmetric and antisymmetric, under the operation of parity.
This is precisely what we find when we examine those figures. In addition,
we also note that both the orbitals have nodes near the positions 
$(\pm a/4,\pm a/4,\pm a/4)$, which correspond to the locations of the two
Li atoms of the unit cell. These nodes are a consequence of the 
orthogonalization of the Wannier functions to the Li 1s-type core Wannier
functions. The localized nature of both the Wannier functions is evident
by their rapid decay as one moves away from the oxygen site. This is a rather
pictorial confirmation also of the ionic character of the crystal.
The noteworthy point is that apart from the presence of the
orthogonality nodes which are indicative of the presence of the 
environment, the qualitative nature of Wannier functions is very similar
to those of any molecular orbitals, thus bringing them close to the intuition
of a chemist.

The simplest qualitative picture of cohesion in an ionic solid is that it is 
accompanied by charge transfer from one set of atoms to another, leading to 
cations and anions at different sites. These oppositely charged ions, in turn,
are held together by the electrostatic attraction between them. For the 
present case of Li$_2$O,
this would imply that the crystal consists of Li$^+$ and O$^{--}$ ions in 
closed-shell configurations. Such a picture can be
verified, e.g. by performing a Mulliken population analysis, also in a 
Bloch-orbital-based approach~\cite{torino1}. However, a 
Wannier-function-based approach provides a more direct and unambiguous view of the
cohesive process.
To this end we first compare the 2s orbital of the free oxygen atom (which is
also plotted in Fig.\ \ref{fig-comp2s111}) with the
2s-type Wannier function of Li$_2$O. For free oxygen atom a basis set
also reported by the Torino group~\cite{torino1} was used, which they obtained
by reoptimizing the outermost exponents of the crystal basis set and by adding
one more sp-type diffuse exponent (0.0867) to it.
As is clear from the figure, the spatial behavior of
the 2s orbital changes only in a rather subtle manner in that the
corresponding Wannier function develop nodes near 
the positions $(\pm a/4,\pm a/4,\pm a/4)$  due to their
orthogonalization to the 1s-type Wannier functions of the Li$^{+}$ ions of the
unit cell located there. Apart from these orthogonality nodes, the two orbitals
are remarkably similar.
This is consistent with our intuitive picture in that being a lower valence 
orbital, one should not expect the 2s orbital to be affected appreciably by 
the crystal field.
Therefore, we expect the effects of cohesion to be most evident in the case
of the 2p-type Wannier functions. In the simple ionic picture, the O$^{--}$ ion
of the solid has a filled 2p shell. The presence of two extra electrons on 
oxygen will result in the 2p electrons experiencing 
larger on-site repulsion leading to the charge density associated with
2p-type Wannier functions becoming more diffuse compared to the isolated 
atom. Using the same intuitive reasoning we expect the 2p charge density
in the solid to be more diffuse even compared to the free O$^{-}$.
These points are illustrated in Fig.\ \ref{fig-comp2p111} where we plot
the total 2p charge density associated with the 2p-type Wannier functions  of 
the Li$_2$O lattice along the [111] crystal direction, and compare it with 
the corresponding charge densities of free O atom and free O$^{-}$ ion.
The orbitals of free O$^{-}$ ion were obtained from a basis set which was
obtained by augmenting the free atom
basis set discussed earlier by one more diffuse sp-type exponent of 0.0345.
The total normalized charge densities were obtained by proportionately adding
the contributions 
of the individual charge densities of the three 2p-type orbitals of the
systems under consideration.   
A delocalization trend is certainly obvious in the region close to the
nuclei from the 
decreasing heights
of the peaks as one compares the 2p-charge densities in the order of
free O atom, free O$^-$ ion and embedded O$^{--}$ ion. To quantify the
degree of localization of various orbitals  we computed the
directionally averaged expection values of $r^2 = x^2+y^2+z^2$ operator for
the 2p-type orbitals of free O, O$^-$ and the corresponding Wannier
functions of the Li$_2$O crystal. The directionally averaged values of a given
system were obtained
by averaging over the expectation values of the three 2p-type orbitals
according to $\overline{<r^2>}_{2p} = 
(\sum_{i=1}^3 <r^2>_{2p_{i}})/3$.The obtained values for 
$\overline{<r^2>}_{2p}$ (in  Bohr$^2$) were $1.974$ (free O atom), 
$3.038$ (free O$^-$ ion) and, $3.008$ (Li$_2$O crystal).
As expected, we find that the value of $\overline{<r^2>}_{2p}$ is smallest
for the case of free O atom. However, when we compare the corresponding
values of free O$^-$ with those of the embedded O$^{--}$ of the Li$_2$O
crystal, we arrive at somewhat counter-intuitive result that the size of
the 2p-type orbital of the embedded O$^{--}$ is a little smaller than
the corresponding orbital of free O$^-$. To account for finite-basis-set
effects as well as the nonuniqueness of Wannier functions, we can 
perhaps say that the 2p-type orbitals of free O$^-$ and embedded 
O$^{--}$ are approximately equal. 
One possible objection
against this finding is that our basis set for the solid-state
calculations, with the most diffuse sp-type exponent of only 
0.15~\cite{torino1}, is not diffuse enough to describe a doubly charged 
anion like O$^{--}$. However, we do not suspect this to be the reason behind
our finding because, as discussed earlier, our Wannier functions 
contain basis functions not only of the atom that they are centered upon,
but also the basis functions of atoms contained in up to third-nearest
neighbour unit cells of the reference cell. Thus these moderately diffuse
exponents located on the neighboring atoms should be able to describe
the long-range behavior of the Wannier functions of the reference atoms.
Moreover, since the solid state basis set has correctly predicted the
lattice constant of Li$_2$O, we expect it to be a reasonably complete set.
Our explanation of the finding, that the size of embedded O$^{--}$ ion of
the Li$_2$O crystal is essentially the same as that of 
the singly charged free O$^{-}$, is based upon a competition between the
on-site repulsion and the crystal-field effects (Pauli repulsion of
surrounding atoms). 
In our opinion, in the crystalline phase, the on-site repulsion which 
certainly has a delocalizing influence, is overpowered by the crystal
field  leading to a stable O$^{--}$ ion whose size is comparable to that 
of the free O$^{-}$ ion.

\section{CONCLUSIONS}
\label{conclusion}
 In this paper we extended the application of a Wannier-function-based
Hartree-Fock approach developed earlier~\cite{shukla1,shukla2} to alkali 
oxides, Na$_2$O and Li$_2$O. State-of-the-art basis sets containing rather
diffuse s- and p-type functions were used and quantities such as the
equilibrium lattice constant, bulk modulus and cohesive energy per unit
cell were computed. We obtained excellent agreement with the results
of a previous Hartree-Fock study carried out by the Torino group~\cite{torino1}
using their Bloch-orbital-based CRYSTAL 
program~\cite{crystalprog}. In addition, a detailed pictorial view of the cohesive process is
provided by the Wannier functions of our work which comes close to the 
intuitive understanding of a chemist. 

As far as the agreement with experimental data is concerned, it is excellent
for the lattice constant. For the bulk modulus agreement is acceptable. 
As expected, the most severe deviation is evident in case of the cohesive energies
of the two compounds where disagreement between Hartree-Fock theory 
and experiment is in the range of 30\%---40\%.
This points to the importance of correlation effects,
and we consider it as an advantage of our
Wannier-function-based Hartree-Fock approach that it provides a natural
starting-point for improvement by means of local correlation methods~\cite{corr}.
Efforts along this 
direction are presently underway in our group, and the results will be 
presented in a future publication.

\pagebreak
%
\begin{figure}
\caption{Free atom 2s orbital of oxygen and 2s-type crystal Wannier function
of Li$_2$O localized on O$^{--}$ plotted along the [111] direction.
}
\label{fig-comp2s111} 
\end{figure}
\begin{figure}
\caption{2p-type crystal Wannier function of Li$_2$O
localized on O$^{--}$ plotted along the [111] direction.
}
\label{fig-2p111} 
\end{figure}
\begin{figure}
\caption{Charge density associated with the 2p-type Wannier functions 
of Li$_2$O localized 
on O$^{--}$, compared to the 2p charge densities of the free oxygen atom and free O$^{-}$. 
All the charge densities are plotted along the [111] direction and are
normalized to unity.
}
\label{fig-comp2p111} 
\end{figure}
\begin{table}  
 \protect\caption{ Equilibrium lattice constants (in \AA), bulk moduli 
(in GPa), and cohesive energies 
(in atomic units) for Na$_2$O  and Li$_2$O, obtained using our approach and 
reported by the Torino group\protect\cite{torino1}. Relevant experimental data
are also given for comparison.}
  \begin{tabular}{|l|lll|} \hline  \hline   
 Quantity  & Method    & \multicolumn{2}{c|}{System}  \\ 
           &           &  Li$_2$O     &   Na$_2$O   \\ \hline \hline 
           & This Work & 4.570        & 5.481        \\
Lattice Constant
           & Torino   & 4.573        & 5.484             \\
           & Exp       & 4.573$^a$   & 5.55$^b$      \\ 
           &           &             & (5.49)  \\ \hline
           & This Work & 94.6        & 61.1             \\
Bulk Modulus
           & Torino   & 92.6         & 58.4               \\
           & Exp       & 89$^a$      & ---            \\ \hline
           & This Work & -0.3008      & -0.1883              \\
Cohesive Energy
           & Torino   & -0.3005      & -0.1893             \\
           & Exp       & -0.4491$^c$       
                                      & -0.3321$^c$ 
                  \\ \hline \hline   
   \end{tabular}                      
  \label{tab-eqprop}    
$^a$ Ref.~\cite{expali2o}. \\
$^b$ Room temperature value from ref.~\cite{expana2o}. The number in 
parentheses is the extrapolated zero-temperature value
suggested by the Torino group~\cite{torino1}. For a discussion see text. \\
$^c$ Ref.~\cite{beexp}.  
\end{table}  
\centerline{\psfig{figure=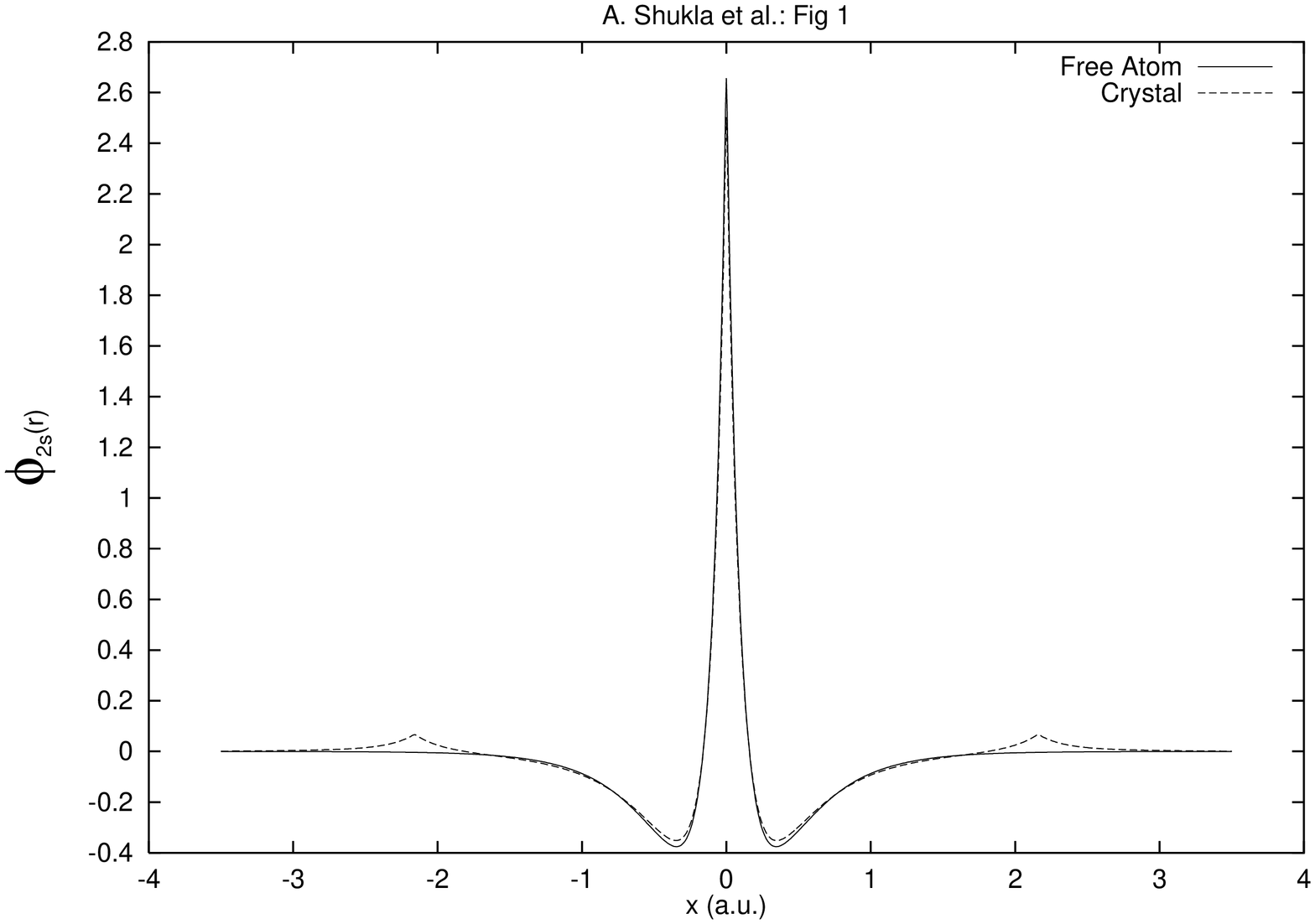,width=13cm,angle=-90}}
\centerline{\psfig{figure=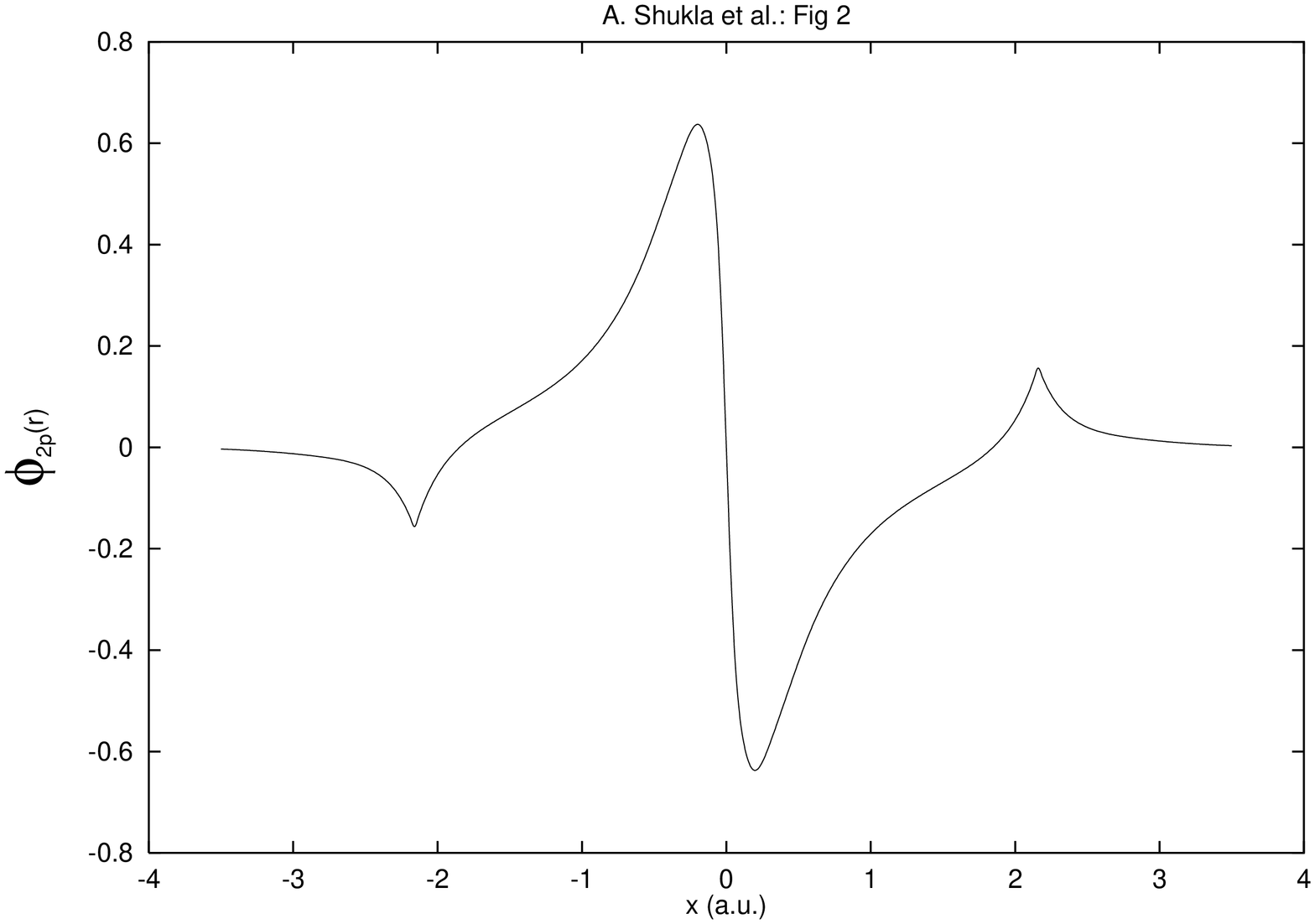,width=13cm,angle=-90}}
\centerline{\psfig{figure=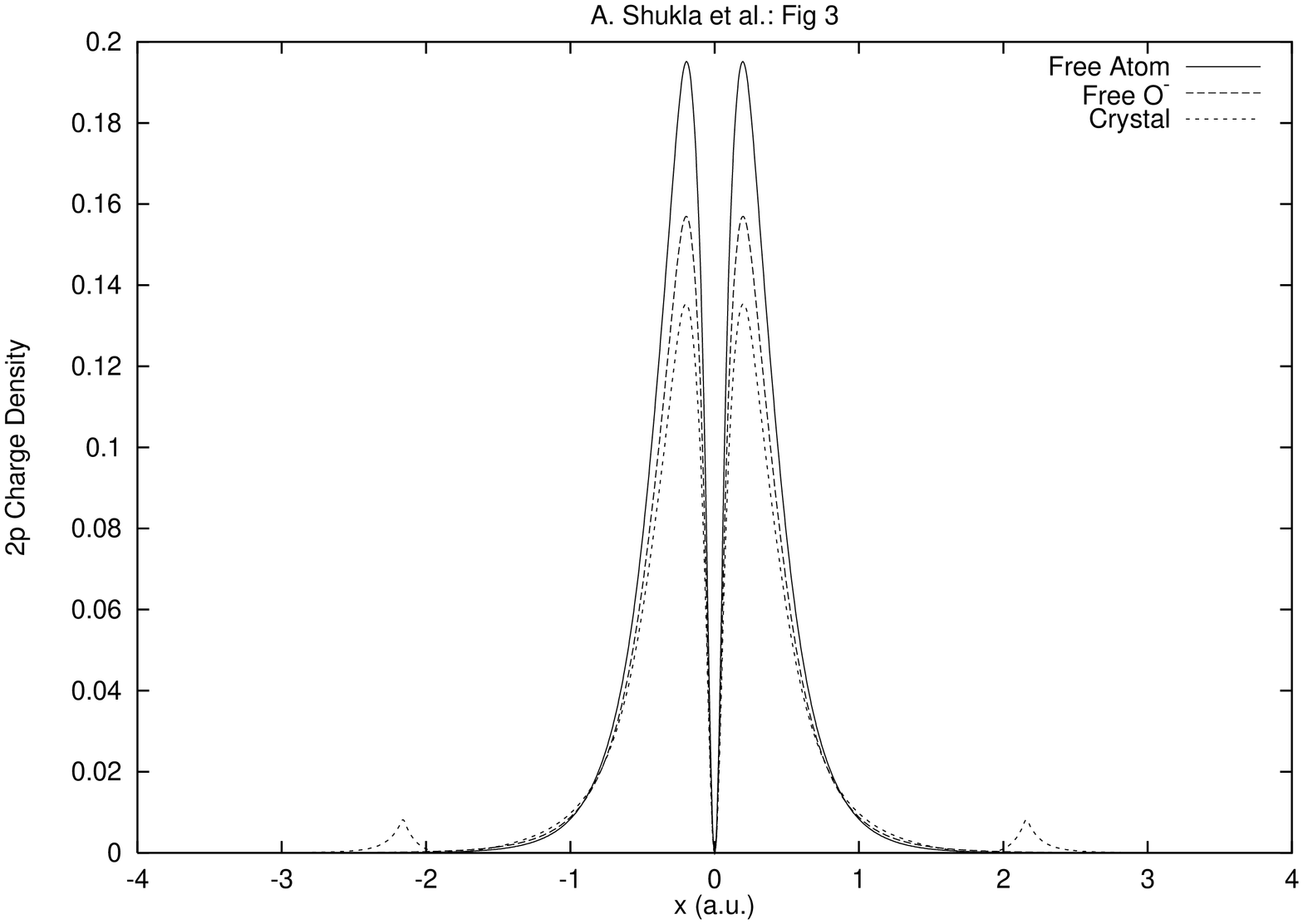,width=13cm,angle=-90}}
\end{document}